\documentclass[letterpaper, 10pt, conference]{ieeeconf}      

\usepackage{amsmath}

\usepackage{amssymb}
\usepackage[utf8]{inputenc}
\usepackage{algorithm}
\usepackage{algorithmic}
\usepackage{graphicx}

\usepackage{subfig}

\usepackage{graphicx}
\DeclareMathOperator*{\argmin}{arg\,min} 
\DeclareMathOperator*{\argmax}{arg\,max} 
\usepackage{centernot}
\usepackage{lipsum}
\usepackage{hyperref}
\usepackage{blindtext}
\usepackage{scrextend}
\usepackage{stackengine}
\usepackage{tikz}

\IEEEoverridecommandlockouts                              

\title{\LARGE \bf
Optimizing 
Consensus-based Multi-target Tracking \\
with Multiagent Rollout Control Policies

}

\author{Tianqi Li$^{1}$, Lucas W. Krakow$^{2}$ and Swaminathan Gopalswamy$^{1}$
\thanks{$^{1}$Tianqi Li and Swaminathan Gopalswamy are with the Department of Mechanical Engineering, 
        Texas A\&M University, College Station, TX 77840, USA
        {\tt\small \{xmcx731, sgopalswamy\}@tamu.edu}}%
\thanks{$^{2}$ Lucas W. Krakow is with the Bush Combat Development Complex (Texas A\&M)
            Bryan, TX 77807, USA
        {\tt\small lwkrakow@tamu.edu}}
\thanks{This work has been submitted to the IEEE for possible publication. Copyright may be transferred without notice, after which this version may no longer be accessible.}
}
\specialpapernotice{(Invited Paper)}

\begin{document}

\maketitle
\thispagestyle{empty}
\pagestyle{plain}

\begin{abstract}

This paper considers a multiagent, connected, robotic fleet where the primary functionality of the agents is sensing. 
A distributed multi-sensor control strategy maximizes the value of the collective sensing capability of the fleet, using an information-driven approach. 
Each agent individually performs sensor processing (Kalman Filtering and Joint Probabilistic Data Association) to identify trajectories (and associated distributions). 
Using communications with its neighbors the agents enhance the prediction of the trajectories using a \textit{Consensus of Information} approach that iteratively calculates the Kullback-Leibler average of trajectory distributions, enabling the calculation of the value of the collective information for the fleet. 
The dynamics of the agents, the evolution of the identified trajectories for each agent, and the dynamics of individual observed objects are captured as a Partially Observable Markov Decision Process (POMDP). 
Using this POMDP and applying rollout with receding horizon control, an optimized non-myopic control policy that maximizes the collective fleet information value is synthesized. 
Simulations are performed for a scenario with three heterogeneous UAVs performing coordinated target tracking that illustrate the proposed methodology and compare the centralized approach with a contemporary sequential multiagent distributed decision technique.
\end{abstract}


\section{Introduction}

In the paradigm of autonomous systems, sensing is regarded as the primitive which transforms information of the physical world into the system as input for example action selection or planning. 
Interactions between sensing and other primitives support the tasks which employing sensor information in world modeling, stimulus of controller and knowledge representation.
Research on control strategies with considerations centered around information and sensors, including but not limited to target tracking, searching, world mapping, is a new and fast-growing focus.
Information-driven \cite{zhao2002information, 
bellini2014information}, or sensor-driven control \cite{paull2012sensor}, studies the problem of optimizing a system's configuration and resource allocations to optimize or improve sensing performance.

There are two main schemes in data processing of multiagent systems, \emph{centralized} and \emph{distributed}.
In a centralized architecture, every agent must send its information (observation, processed message, control information etc.) to a central control node either directly or by some multi-hop relay. 
This configuration can instigate challenges with respect to the growth of the system's  number of agents and this is compounded when the agents are mobile:
the central node has bandwidth linear to the number of agents;
the topological change in the communication graph caused by moving agents makes it hard for the message be passed by all nodes in the network \cite{xiao2005scheme}.
The distributed approach means there is no fixed information aggregation component central to the operation of the system, and the communication is peer to peer \cite{siciliano2016springer}.
In applications like sensor fusion, distributed processing may negatively impact the system's overall performance when compared to a centralized method, its advantages are realized in sensor resource management and the robustness of the system, e.g., communications.
Decentralized sensor fusion can be more adaptable in applications like  sensor deployment for maximal coverage \cite{cortes2004coverage}, where sensors that require local decisions already possess the needed local information such as target tracking estimates.
For the aspect of communication, constraint in bandwidth makes the centralized sensor fusion infeasible to be for large system implementation \cite{luo2005universal}.

The trend of applying decentralized algorithms is arising in the information-driven tasks, leveraging the aforementioned advantages.
\cite{olfati2007distributed} establishes the link of distributed target tracking and information-driven mobility, as a byproduct the behavior of flocking is observed in single target tracking by a mobile multisensor network.
\cite{ristic2020intermittent} presents the coordinated intermittent searching of robot team in a restricted area with max-consensus algorithm on the information to picture the occupancy map, in which mobile sensor formations are propelled by average consensus on decisions.

However, an optimal solution for such coordinated heterogeneous multiagent planning poses a computational challenge.
\cite{zhou2011multirobot} introduces the sensor planning based on the estimation of a single target with the range-bearing sensor model and motion constraints, and studies such a strategy in single and (heterogeneous) multisensor scenarios.
With motion constraints on the robots (maximum speed and minimum distance to the target), such non-convex constrained optimization problems are NP-hard in general.
The work in \cite{ramachandran2020resilience} strives to maximize the FoV of the sensing team by topological reconfiguration of the robot network under sensor deterioration. 
The solution to the proposed reconfiguration problem equates to a circle packing problem in 2-dimension space, which is shown to be NP-complete \cite{demaine2010circle}.
It is clear that optimality is difficult to achieve, thus, efficient suboptimal solutions are of great value for implementations.
Recently a multiagent version of rollout policy was published \cite{bertsekas2019multiagent}, which takes augments the traditional rollout approach through sequential optimization. 
In a multiagent system containing $n$ agents with $\mathbb{R}^s$ as the action domain of each agent, the joint optimization requires the computation of order $O(s^n)$. 
However, when performing \emph{sequential} optimization with respect to the agents, rollout reduces the computation load to the order of $O(sn)$.

This paper focuses the coordination of mobile multi-sensor system in the task of target tracking.
Different from work in \cite{zhou2011multirobot}, scenario of tracking multiple targets is studied for a more generic version of distributed target tracking.
For the simplicity of the problem, agents are considered as UAVs for freedom of movement and alleviate  sensor-to-target and sensor-to-sensor distance constraints. 
Since future targets positions are stochastic, a non-myopic behavior is of great interest in the planning. This is achieved through the application of a receding horizon when computing rollout policies.
We formulate a POMDP framework and apply a sequential rollout method in a consensus-based distributed target tracking task to solve an information-driven coordinated planning with a heterogeneous multiagent system.
With simulation designed in target tracking, the performance of tracking in rollout algorithm shows
\begin{enumerate}
    \item The non-myopic behavior extracted from the multiagent rollout benefits the tracking performance for the fleet of mobile sensors 
    \item The computation advantages of the sequential optimization in multiagent rollout policy can be realized with minimal target tracking degradation.
\end{enumerate}

This paper is organized in the following manner.
Section \ref{section_problem_statement} describes fundamental setup of distributed target tracking and the objective of the proposed problem.
Section \ref{section_method} details the proposed non-myopic multisensor planning in a framework of POMDP.
Section \ref{section_result} presents the simulation results which demonstrates the coordination of sensors in target tracking task.
Insights of this study are recapped and future work is proposed in Section \ref{section_conclusion}.

\section{Problem Statement}
\label{section_problem_statement}

We consider a heterogeneous mobile multi-sensor system in a task of target tracking for our specific control problem.
Each \emph{agent} is equipped with sensors, actuators and a communication device for the three main functions: sensing, mobility and communicating with peer agents.
This multi-sensor system has total $n$ agents and each agent has a unique label $i \in [n]$.
Define the state of the single agent $i$ as $s^i \in \mathcal{S}^i$, where $s^i$ contains the pose and velocity of the agent in horizontal plane $s^i\in \mathcal{S}^i, s^i = (x, y, \theta, v_x, v_y)$.
We assume a holonomic model for the dynamics of each agent where the control inputs directly alter agent's velocity; the control domain $\mathcal{U}^i$ is the velocity of the agent, such that $u^i\in \mathcal{U}^i, u^i = (v_x, v_y)$ and the azimuth in pose is consistent with velocity, i.e. $\theta = arctan(v_y/v_x)$.
An agent $i$ has the following attributes:
\begin{itemize}
\raggedright
    \item Mobility:  
    For agent $i$, in discrete time space, we have state $s_k^i \in \mathcal{S}^i$ and control $u_k^i \in \mathcal{U}^i$, the sensor state transition function is $ s^i_{k+1} = f_i(s^i_k, u_k^i)$ .
    \item Sensing: Each sensor has a limited field of view (FoV) and an associated \emph{sensing quality} .
    For a quad-copter UAV agent, we define the FoV of agent $i$ by variable $\phi_k^i(s_k^i)$ parameterized by agent state. 
    For sensing quality, a scalar $\alpha_i$ is defined in the measurement uncertainty matrix $R$, detailed in \eqref{observation-model}.
    \item Communication: An agent communicates to other within its neighborhood defined by the distance of $d_c$. For simplicity,  we assume this threshold is universal across agents.
\end{itemize}

The multisensor network is comprised of $n$ agents described above.  The network can then be represented as an undirected proximity graph $\mathcal{G} = (V(\mathcal{G}), E(\mathcal{G}))$, let vertex $v \in V(\mathcal{G})$ represents an agent in graph and every edge $e \in E(\mathcal{G})$ represents communication link between two agents, i.e. $\forall \{i, j\} \in V(\mathcal{G}), ||s^i, s^j|| \leq d_c$.
Define the \emph{neighbor} of agent $i\in V(\mathcal{G})$ as the set of graph vertices $N_\mathcal{G}(i) \in V(\mathcal{G}), \text{s.t. } \forall j \in N_\mathcal{G}(i), \{i, j\} \in E(\mathcal{G})$, and we denote the degree of agent $d(i) = |N_\mathcal{G}(i)|$.

\subsection{Distributed Target Tracking}
\label{tracking}

In the multi-target tracking (MTT) problem, at each time step $k$, an agent $i$ generates a set of measurements deemed as an observation, $Z_k^i = \{z_{k, 1}^i, z_{k, 2}^i..., z_{k, m^i_k}^i \}$ with cardinality $|Z_k^i| = m^i_k$. 
Given a target state $\mathbf{x} \in \chi$ which contains position and velocity variables in a 2-dimensional Cartesian coordinate system, and a measurement, $z \in \mathcal{O}$, generated from a target is taken to be the target's position.

A linear assumption of the dynamic of target is
\begin{equation}
    \textbf{x}_{k+1} = F_k\textbf{x}_k + w_k, ~w_k \sim \mathcal{N}(0, Q_k)
    \label{target-dynamic}
\end{equation}
with $w_k$ represents the motion disturbance in a normal distribution, $F_k$ is the nearly constant velocity motion model, and observation law is 
\begin{equation}
    z_k = H_k\textbf{x}_k + v_k, ~v_k \sim \mathcal{N}(0, R_k)
    \label{target-obs}
\end{equation}
with $v_k$ represents the measurement noise. 
The common \textit{range-bearing sensor} model is considered in our paper, which is widely studied and applied \cite{anderson2012optimal}.
Specifically, for agent $i$ with state $s^i_k$, let $r_k$ and $\rho_k$ denote the estimated range and bearing of a mobile target, $r_k = max(||s^{i, pos}_k - \textbf{x}^{t, pos}_k||, r_0)$, $\rho_k$ is the angular measurement between the sensor and target,
the observation covariance matrix $R_k^i(s^i_k, \textbf{x}^t_k)$ for target $t$ is 
\begin{equation}
\label{observation-model}
R_k^i(s^i_k, \textbf{x}^t_k) = \alpha_i G(\rho_k)\begin{bmatrix}
0.1\times r_k & 0\\
0 & 0.1\pi \times r_k
\end{bmatrix}G(\rho_k)^T
\end{equation}
where $G(\rho)$ is the rotation matrix of angle $\rho$, and sensing quality factor $\alpha_i$ is a scalar in this uncertainty matrix.
This imposes the spatially varying measurement error \cite{krakow2018simultaneous}. 
To avert computational issues inside the Kalman filter updates, we apply $r_0$ as the minimal effective range threshold.

Based on the principle of distributed target tracking in \cite{sandell2008distributed}, the systematic estimation is twofold: i) single agent local estimation and prediction; ii) track fusion over agents.
For a single agent, the estimation includes the measurement-to-track assignment and estimation update given measurement.
One method of this measurement-to-track assignment is the joint probabilistic data association filter (JPDAF) \cite{fortmann1983sonar}.
JPDA algorithm handles the assignments based on conditional probabilities calculated with respect to predicted target locations and the received measurements.
Since the target may be outside of sensor FoV or simply undetected, we also employ \textit{M-out-of-N} track maintenance logic for both track confirmation and deletion. 
A track is confirmed/deleted if there are more than \textit{M} detections (or missed detections, respectively) in the latest \textit{N} consecutive observations.

Once the local tracking step (JPDAF) is complete, messages containing tracking information from an agent's neighbors are received and processed to update local track in the consensus period.
As is common in distributed sensor fusion, we follow the average consensus scheme for this  step.
However, instead of sending the measurements, we perform \textit{consensus of information} (CI) which directly processes the tracking information. 
For example, \cite{battistelli2014consensus} utilizes the information filter to make average consensus of track information, and the result is proven to be the Kullback–Leibler average (KLA) of the initial local distributions \cite{battistelli2014kullback}.
Denote $\Theta^i_k$ as the set of local tracks of agent $i$. 
In a Kalman filter, a pair consisting of the mean vector and covariance matrix ($\textbf{x}^t_{i, k|k}, \textbf{P}^t_{i, k|k}$) is the standard representation of a track $t \in \Theta^i_k$.
The same track can be represented in information filter by the pair comprised of an information matrix 
$\mathbf{\Omega}^t_{i, k|k} = {\textbf{P}^t_{i, k|k}}^{-1}$ and an information vector $\textbf{q}^t_{i, k|k} = \mathbf{\Omega}^t_{i, k|k}\textbf{x}^t_{i, k|k}$.

As with any consensus algorithm, multiple updates must be done for the information to converge across the network.
We denote the total number of consensus update steps as $L$, which take place in order to accurately proliferate the JPDA tracking information to all agents of the network. 
These $L$ updates are taken over each discrete epoch, e.g., from $k$ to $k+1$. 
At each consensus update $l = 1,2, \ldots, L$ the average consensus step for agent $i$ is
\begin{subequations}
\label{consensus}
\begin{align}
\mathbf{\Omega}^t_{i, k|k}(l+1) =
& \frac{1}{1 + d(i)}\sum_{j \in N_G(i)\cup \{i\}}\mathbf{\Omega}^t_{j, k|k}(l)  \label{consensus:1}\\
\mathbf{q}^t_{i, k|k}(l+1) =
& \frac{1}{1 + d(i)}\sum_{j \in N_G(i)\cup \{i\}}\mathbf{q}^t_{j, k|k}(l)   \label{consensus:2}
\end{align}
\end{subequations}
Given the fixed number of consensus steps, the output of CI is $(\mathbf{\Omega}^t_{i, k|k}(L), \mathbf{x}^t_{i, k|k}(L))$. 

\subsection{Information-driven multiagent Coordination Control}
\label{information-driven}
The multisensor system described above is information-driven, i.e. the control of mobile sensors is to obtain over all more information of targets by observation and distributed target tracking.
Firstly a good choice of information utility is essential to picture the objective of the system in a mathematical formation.
The existing choices includes entropy \cite{zhao2002information}
, Fisher Information matrix \cite{olfati2007distributed}, the covariance matrix in Gaussian \cite{zhou2011multirobot,ramachandran2020resilience, krakow2018simultaneous} and other sensor model-based heuristics \cite{chu2002scalable}.
In our problem, the trace of covariance matrix of targets is chosen for its perceptual intuition and ease of computation in the Gaussian based data association.
To further simplify the computation, since the information matrix $\mathbf{\Omega}^t_{i, k|k}$ is the inverse of covariance matrix $\textbf{P}^t_{i, k|k}$, the trace of information matrix is applied as the utility of information.
In our distributed multisensor system, the utility of information of agent $i$ in time $k$ is defined as 
\begin{equation}
    \label{info-utility}
    \Phi^i(k) = \sum_{t \in \Theta^i_k} tr(\mathbf{\Omega}^t_{i, k|k}(L))
\end{equation}
Notice that this utility contains the process of distributed MTT in Section \ref{tracking}, term $\mathbf{\Omega}^t_{i, k|k}(L)$ is defined as the output of consensus step of certain track $t$ by \eqref{consensus:1}.

\subsection{Semantic Map}
The study of sensing is dependent on the environment, and in our mobile multisensor system the physical environment has impact on sensibility of agents.
The semantic meaning of areas and objects helps in understanding the environment and provides guidance on robot decision making.
Following \cite{zheng2018pixels}, define the world $W =(T, X, Y)$ with a topological graph $T$ to represent components of physical object set $X$ and semantic element set $Y$.
In our problem, geological factor affects the quality of detection like tree shadows and ground reflection, by utilizing the semantic map, the movement of targets into occlusion area will be predicted based on Kalman filter. 
Such semantic information helps picturing the future observations based on current target belief.
Given a set of observation $Z$, semantic map functions as $W(Z)$ which returns the observation outside the occlusion area.


\section{Method}
\label{section_method}

\subsection{A POMDP Model}
A Partial Observable Markov Decision Process (POMDP) model is formulated in \cite{krakow2018simultaneous}, which describes a non-myopic sensory centralized multi agent motion planning with the task of target tracking. 
This POMDP model is extended to the multi agent scenario with distributed planning in this paper.
Define a POMDP model as a tuple of $\mathcal{P} = (\mathcal{X, U, T, R, O})$.

\textbf{State} $\mathcal{X}$: The state of POMDP contains the state of agents $\mathcal{S}$, state of targets $\chi$ and state of filter $\mathcal{F}$.
The state of all $n$ agents is defined as $\mathcal{S} = \mathcal{S}^1 \times \mathcal{S}^2 ... \times \mathcal{S}^n$.
Together with agents states $s_k \in \mathcal{S}$, the overall FoV of these agents is $\phi_k(s_k) = \phi^1_k(s_k^1) \times \phi^2_k(s_k^2) ... \times \phi^n_k(s_k^n)$, which is regarded as a variable parameterized by agents state.
Targets state $\chi$ contains all position and velocities of current targets.
The filter state $\mathcal{F}_k = \{ (\textbf{x}_{i, k|k}^t, \textbf{P}_{i, k|k}^t) | \forall i \in [n], \forall t \in \Theta^i_k\}$, which are all agents maintained tracks represented by Gaussian distributions with posterior means $\textbf{x}_{i, k|k}^t$ and posterior covariance matrices $\textbf{P}_{i, k|k}^t$.
The POMDP state is summarized as $x_k = (s_k,\chi_k, \mathcal{F}_k)$.

\textbf{Action} $\mathcal{U}$: The action domain is the overall action to the multisensor system, with the control domain of each sensor defined by the 2-d velocity in Section \ref{section_problem_statement}.
Such control domain has constraint from both the physical world and the agents' kinematic constraints.
Assume the maximum velocity is set as $v_max$ for all agents $i\in [n]$.
Thus at time $k$, the control domain $\mathcal{U}^i_k$ of a single agent $i$ is implicit with its state $s_k^i$.
The joint action domain is then defined as $\mathcal{U}_k = \mathcal{U}_k^1 \times ... \times \mathcal{U}_k^n$.

\textbf{Observation and Observation Law} $\mathcal{O}$:
Observation is generated by sensors equipped on agents through perception algorithm, here we assume that an ideal perceptual component outputs observations with no false alarms in areas without occlusion in $W$.
Under Gaussian noise assumptions, the observation law of a target is presented by \eqref{target-obs} with the dynamic sensor model defined as \eqref{observation-model}. 
From a system perspective, the joint observation domain of all sensors is defined as $\mathcal{O}$, $Z_k \in \mathcal{O}, Z_k = \{Z_k^i|i\in [n]\}$.
For agent $i$, its observation $Z_k^i$ is constrained by the semantic map $W$ and its FoV $\phi^i_k$.

\textbf{State Transition} $\mathcal{T}$: 
State transition law is defined as the mapping $\mathcal{T}: \mathcal{X} \times \mathcal{U} \xrightarrow[]{} \mathcal{X}$.
In the state $x_k\in \mathcal{X}_k, x_k = (s_k, \chi_k, \mathcal{F}_k),u_k\in\mathcal{U}_k$, the state transition is decomposed into the following:
i) the agents' transitions are completely determined by $ s^i_{k+1} = f_i(s^i_k, u_k^i)$;
ii) target states $\textbf{x}_k^t \in \chi_k$ are a stochastic process \eqref{target-dynamic} with Gaussian additive noise, independent of sensor state $s_k$ and control variable $u_k$.
iii) the filter state $\mathcal{F}_k$ is dependent on both agent state $s_k$, FoV of sensors $\phi_k$ and target state $\chi_k$ and evolves according to the chosen tracking filter equations, JPDA with average consensus.

\textbf{Reward} $\mathcal{R}$: The reward function for single agent is a mapping $\mathcal{X} \times \mathcal{U} \xrightarrow[]{} \mathbb{R}$, with the objective of maximizing the tracking information.
For agent $i$, its tracking set $\Theta^i_k$ can be regarded as the subset of all targets in the system $\Theta^i_k \subseteq \chi_k$.
The standard objective in MTT is to minimize the tracking error  over the system.  
By design, the information utility function defined in \eqref{info-utility} directly reflects our purpose by maximizing the trace of the inverse of uncertainty matrix over all tracks.
In \cite{battistelli2014kullback} it is shown that under the assumption of strongly connectivity in the sensor network, the estimation error of \textit{CI} is asymptotically bounded in mean square error. 
Let $\Tilde{\Phi}(\cdot)$ denote the \emph{converged estimate}. 
We assume given $L > L_0, \forall i \in [n], |\Phi^i(k) - \Tilde{\Phi}(k)| < \epsilon$, with $\epsilon$ sufficiently small this error can be ignored. 
This implies that the track information of each agent is close enough after sufficient steps of consensus that each agent may take its local estimate to \emph{be} the converged estimate of the system, i.e.,  $\Phi^i(k) \approx \Tilde{\Phi}(k)$. Then the one-step reward for the system is 
\begin{equation}
    \label{reward}
    \mathcal{R}(x_k, u_k) = E_{v_k, w_k}[\Tilde{\Phi}(k+1)|x_k, u_k]
\end{equation}


This POMDP contains two challenging points for an optimal solution: first the action domain expands exponentially with number of agents which leads the computational challenge; second the state transition $\mathcal{T}$ is a continuous stochastic process.

\subsection{Belief-State MDP}

Different from POMDP, the Markov Decision Process (MDP) is fully observable in the state as well as state transition model, which makes it possible to calculate optimal policy given a state and Q-function.
To solve the POMDP, it can be transferred to an MDP problem with \textit{belief state}, $b_k$: in state $x_k = (s_k, \chi_k, \mathcal{F}_k)$ definition, the unobservable part is the target state $\chi_k$ which is represented by the posterior distribution conditioned on the history of actions and observations, i.e. $b_k(x) = P_{\chi_k}(x| Z_0, ..., Z_k; u_0, ..., u_{k-1}) \in B(\mathcal{X})$, here $B(\mathcal{X})$ represents the domain of belief state.
Then the one-step belief-state reward is defined as
\begin{equation}
    \label{belief-reward}
    \Tilde{\mathcal{R}}(b_k, u_k) = \int \mathcal{R}(x, u_k)b_k(x) dx
\end{equation}
replacing \eqref{reward}.
Under the assumption of perfect data association and a linear Gaussian system, a target’s belief state can be represented by a Gaussian distribution such that for target $\chi^i \in \chi_k,$ $b_k^{\chi^i} \sim (\textbf{x}_{i, k|k}^{\chi^i}, \textbf{P}_{i, k|k}^{\chi^i})$. 
Based on this we take the mean and covariance derived from the JPDAF as a sufficient representation of the belief state for the belief-state MDP reward calculations.
The belief state update is then captured using the defined state transition for the fully observable portions, i.e., agent state and the target belief-state update is 
$f_b: B(\mathcal{X}) \times \mathcal{U} \xrightarrow[]{} B(\mathcal{X})$, approximated by the filter state update.
With belief state $b_0$, the cumulative belief reward over $N$ decision stages (horizon) is 
\begin{equation}
    \label{utility}
    V_N(b_0) = E\{\sum_{k=0}^{N} \Tilde{\mathcal{R}}(b_k, u_k) | b_0\}
\end{equation}

The \emph{policy} in belief-state MDP is a mapping $\mu: B(\mathcal{X}) \xrightarrow[]{} \mathcal{U}$ which provides the action given belief state.
In a belief-state MDP, an optimal policy generates the action to maximize the cumulative reward in \eqref{utility}.
Following Bellman's principle, the optimal objective function is interpreted as $V_N^*(b_0) = \max_u \Tilde{\mathcal{R}}(b_0, u) + E[V^*_{N-1}(b_1)|b_0, u]$.
Thus, we define a Q-value function in the fixed horizon as $Q_N: B(\mathcal{X}) \times \mathcal{U} \xrightarrow[]{} \mathbb{R}$, specifically
\begin{equation}
    \label{Q-value}
    Q_N(b_0, u) = \Tilde{R}(b_0, u) + E[V^*_{N-1}(b_1)|b_0, u]
\end{equation}
If N is fixed as constant, this problem becomes \textit{receding horizon control} and the optimal policy $\pi^*$ is defined
\begin{subequations}
\label{receding}
\begin{align}
Q_N(b, u) 
& = \Tilde{R}(b, u) + E[V^*_{N-1}(b')|b, u] \label{receding:1}\\
\pi^*(b) &= \argmax_{u\in \mathcal{U}} Q_N(b, u)   \label{receding:2}
\end{align}
\end{subequations}
where state $b'$ in \eqref{receding:1} is the random belief state obtained from state $b$ after taking action $u$.

\subsection{Multiagent Rollout Policy}

The \textit{rollout} algorithm is a suboptimal solution in decision making which implements a sequential optimization method over a finite time horizon \cite{bertsekas2010rollout}.
Given a simple \textit{base policy} (typically an easy to compute heuristic), the rollout policy is guaranteed to improve upon the base policy.
A multiagent version of rollout is introduced recently by Bertsekas in \cite{bertsekas2019multiagent} which maintains a policy improvement property under a sequential decision making with respect to agents.
In standard one-step lookahead rollout algorithm, from initial state $x_0 \in \mathcal{X}$ a trajectory is generated $\{x_0, \Tilde{u}_0, x_1, ..., x_{N-1}, \Tilde{u}_{N-1}, x_N\}$ with one-step reward function $g_k(x_k, u_k)$ and reward-to-go function $J_{k, \pi}(x_k)$, and a sequence of actions called policy $\pi = \{\Tilde{u}_0, ..., \Tilde{u}_{N-1}\}$ is determined by one-step lookahead maximization
\begin{equation}
    \Tilde{u}_k = \argmax_{u_k} E\{g_k(x_k, u_k) + J_{k+1, \pi}(x_{k+1})\}
    \label{classic-rollout}
\end{equation}
Similarly, the belief-state MDP starts with initial belief $b_0$ and solves such one-step lookahead minimization as \eqref{classic-rollout}. 
Propagating the  the belief into the future is computationally expensive and the accuracy of the estimates decreases with length of the time.
Thus one-step rollout that we implement here focuses on solving the immediate action $\pi^*(b_0)$, which is based on an approximate expected reward-to-go (ERTG) of $E[V^*_{N-1}(b')|b, u]$ in \eqref{receding:1}, and the policy of future beliefs is base policy.



\subsubsection{Base Policy} The base policy of our multi-sensor system is decomposed into individual sensors.
Denote the base policy of the described system as $\Bar{\mu} = \{\Bar{\mu}^1, ..., \Bar{\mu}^n \}$, with each sensor's base policy in a proximal tracking:
the sensor will move towards to the track of largest uncertainty within distance $d_0$, i.e. the track $t^*$ with 
\begin{subequations}
\label{track}
\begin{align}
t^* = \argmin_{t\in \Theta^i_k} tr(\mathbf{\Omega}^t_{i, k|k}) \label{track:1}\\
|s_k^{i, pos} - \textbf{x}^{t^*, pos}_k| \leq d_0^i \label{track:2}
\end{align}
\end{subequations}
For agent $i$, its base policy is 
\begin{equation}
    \label{base-policy}
    \Bar{\mu}^i(s_k^i) = v_0\frac{s_k^{i, pos} - \textbf{x}^{t^*, pos}_k}{|s_k^{i, pos} - \textbf{x}^{t^*, pos}_k|}
\end{equation}
The consideration of proximity in base policy enables the sensor to maintain the target tracking with the simple heuristic of the distance between sensor and its worst local tracking estimate. 
This base policy is specifically motivated by our sensor model with spatially varying measurement error from \eqref{observation-model}.
By moving towards track $t^*$ defined in \eqref{track}, an efficient but myopic single-agent base policy is built.

\subsubsection{Rollout Policy}
The base policy is designed in a distributed single agent's way without coordination among agents. 
The lookahead approach applied by the rollout policy 
implicitly optimizes the agents maneuvers across the system inducing coordinated movements.
The one-step maximization objective function including ERTG term $\Tilde{V}_{N-1}(b) = E[V_{N-1}(b)| b, \Bar{\mu}]$ is
\begin{subequations}
    \label{Q-rollout}
\begin{align}
Q_N(b, u) 
& = \Tilde{R}(b_0, u) + \Tilde{V}_{N-1}(b_0) \label{Q-rollout:1}\\
\pi^*(b_0) &= \argmax_{u\in \mathcal{U}} Q_N(b_0, u)   \label{Q-rollout:2}
\end{align}
\end{subequations}

The optimal solution $\pi^*(b_0) = \{u^1, u^2, ..., u^n\}$ in \eqref{Q-rollout:2} is in a high dimension space $\mathcal{U}$ that brings computational issue in the approach.
In multiagent rollout algorithm, there are two approaches in solving the one-step lookahead optimization: 1) \textit{all-agents-at-once} solves \eqref{Q-rollout:2} in one optimization equation, i.e. returns solution in domain $\mathcal{U}$, which is typically challenged in computation with the growth of dimensionality; 2) \textit{agent-by-agent} initializes the policy $\pi(b_0)$ with all base policy $\Bar{\mu}$ on each agent, doing optimization in $n$ times with $i$th time optimizing the action of agent $i$, solves the joint action policy in a specified sequence of agents.
The algorithm of \textit{agent-by-agent} is described as Algorithm \ref{alg-rollout}.
Such \textit{agent-by-agent} optimization reduces the computation by the trade-off with suboptimality, a brief description of such sequential optimization is in Algorithm \ref{alg-rollout}.

\begin{algorithm}[h!]
\raggedright
\caption{Multi-Sensor Agent-by-agent Rollout Target Tracking}
\label{alg-rollout}
\begin{algorithmic} 
\REQUIRE At time $t = k$, Sensor network graph $\mathcal{G} = (V,E)$, agent state $s_k = (s^1_k, ..., s^n_k) \in \mathcal{S}$, observation set $Z_k = \{Z^i_k, ..., Z^n_k\}$.
\\
\STATE 1. Local sensor update, obtain local tracking set $\Theta^i_k$ for sensor $i$;
\STATE 2. Consensus update with neighbor sensor nodes by \eqref{consensus}, all agent obtains state in POMDP $b_k \in B(\mathcal{X})$;
\STATE 3. Initialize control $u_k$ with base policy $\Bar{\mu}(b_k) = \{\Bar{\mu}^1, ..., \Bar{\mu}^n \}$ by \eqref{base-policy} such that $u_k = \{u_k^i| \forall i \in [n], u_k^i = \Bar{\mu}^i \}$;
\STATE 4. Rollout policy solving one-step optimization
\FOR{$i \in [n]$}
\STATE $u_k^i = \argmax_{u_k^i\in \mathcal{U}^i_k} Q_N(b_0, u_k)$ \\
\STATE update $u_k$
\ENDFOR
\RETURN $u_k$;

\end{algorithmic}

\end{algorithm}

\section{Simulation Result}
\label{section_result}

To validate the aforementioned distributed target tracking rollout policy, a simulation of the multisensor system is constructed described by the following scenario: 3 UAVs carrying camera sensors are deployed within a parking lot perimeter.
The task of UAVs is to monitor the people walking in this parking lot, as we assume the perception algorithm provides the observation data $Z_k^i$ in the frequency of 5 Hz.
To have a close simulation of human's trajectory, step lengths of trajectories obey a Levy walk 
with the speed interval of [1.0, 3.0] m/s.
For this multiagent system, UAVs are set at different altitudes which makes the FoVs $\phi^i_k(\cdot)$ and sensing qualities $\alpha_i$ heterogeneous.
Specifically, UAV 1 has a $20 \times 20$ m square FoV with $\alpha_1=0.1$; UAV 2 has $25\%$ larger side length of FoV and $\alpha_2 = 1.5 \alpha_1$; UAV 3 has $10\%$ larger side length of FoV and $\alpha_3 = 1.2 \alpha_1$.
As the control variable defined in Section \ref{section_problem_statement}, UAVs move on the horizontal planes with fixed altitude by commanding the velocity, with $v_0 = 5$ m/s defined in base policy \eqref{base-policy}, and $d_0^i$ equal to the diagonal of FoV to classify proximal tracks of a certain agent.
We assume that when the sensors move, the FoV maintains orientation consistent with the coordinate system and the same angle through a stabilizing.

\begin{figure}[ht] 

    \centering
    \includegraphics[width=.5\textwidth]{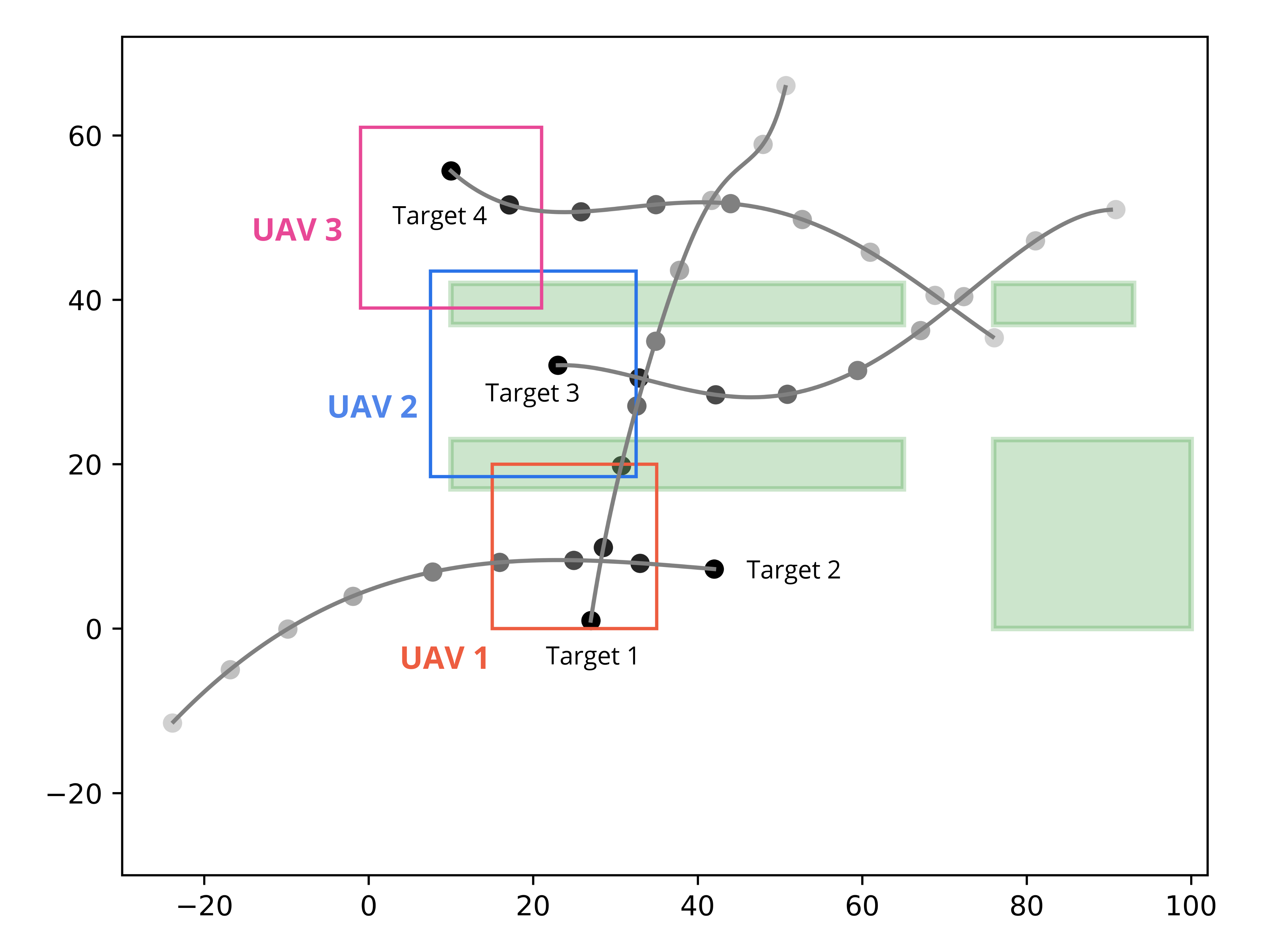}
    \caption{Scenario of 3 UAVs monitoring in a parking lot. The UAV FoV's, denoted as squares, define initial positions. 
    The vegetation areas, solid green blocks, occlude camera detection. 
    The target trajectories are the grey gradient lines from dark (start) to light (end).
    }
    \label{scenario-setup}
\end{figure}

\begin{figure}[htb] 

    \centering
    \includegraphics[width=.47\textwidth]{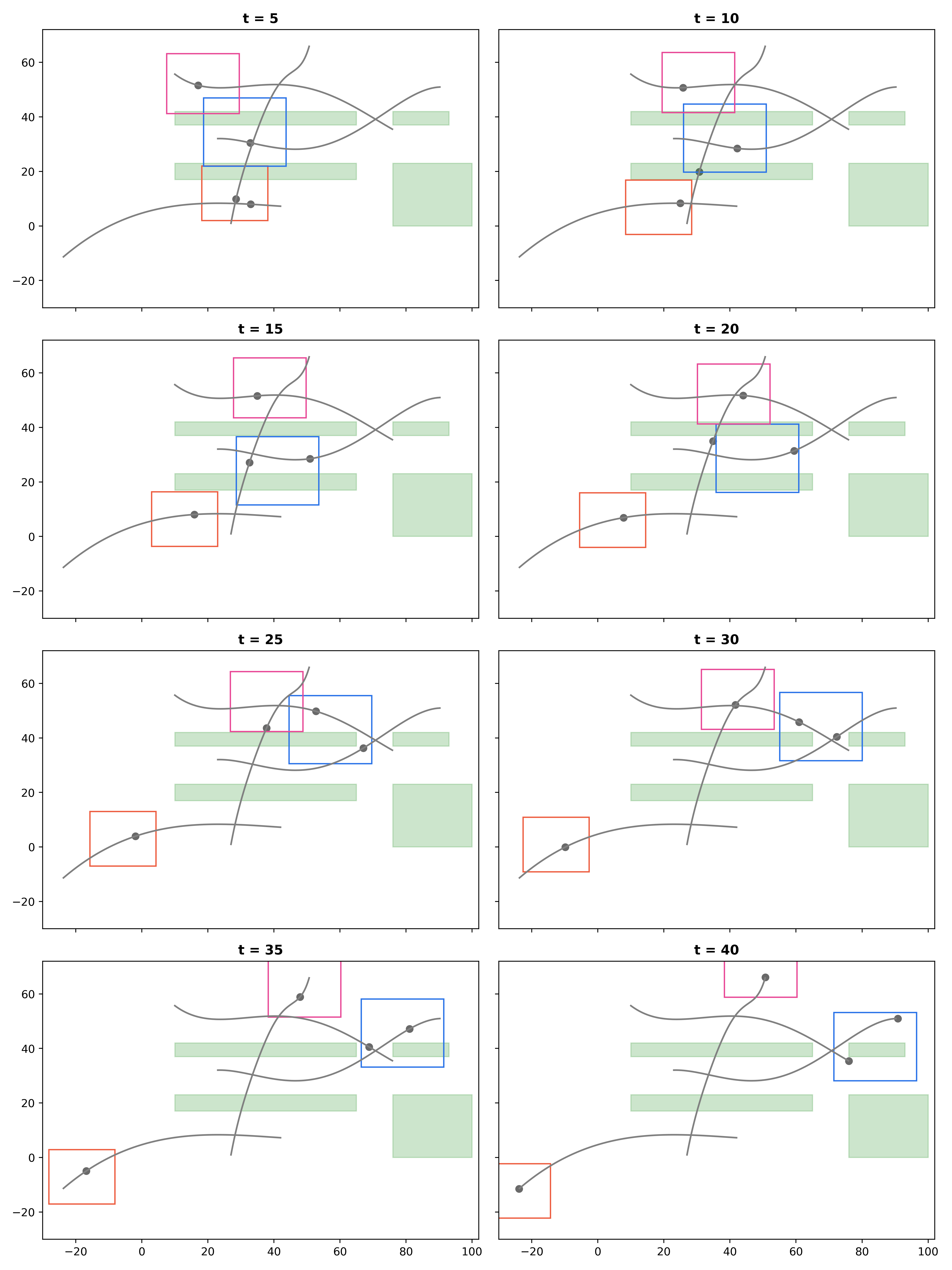}
    \caption{An example tracking result from agent-by-agent rollout, sensors trajectories are dotted lines with corresponding FoVs.
    }
    \label{trajectory}
\end{figure}

The parking lot map is shown in Fig. \ref{scenario-setup}, here the areas of occlusion (tree cover) provides semantic meanings to the system in observation laws defined in POMDP.
Utilizing such semantic information enables the system to plan based on environment and increases the agent coordination.

The control frequency for the system is 1 Hz, which runs rollout policy for all agents.
For a sufficient Q-value approximation in \eqref{Q-rollout}, belief states of target position are represented by 
Monte Carlo sampling \cite{he2006sensor}.
In each control iteration, 50 samples are drawn in the Gaussian belief of targets to formulate the possible trajectories in \textit{N} sec rollout horizon.
Solving of the one-step lookahead optimization of systemwise decision in \eqref{Q-rollout:2} as well as per agent in Algorithm \ref{alg-rollout} is accomplished through numerical optimization, specifically Differential Evolution.

\begin{figure}[h!]
    \centering
    \includegraphics[width = .5\textwidth]{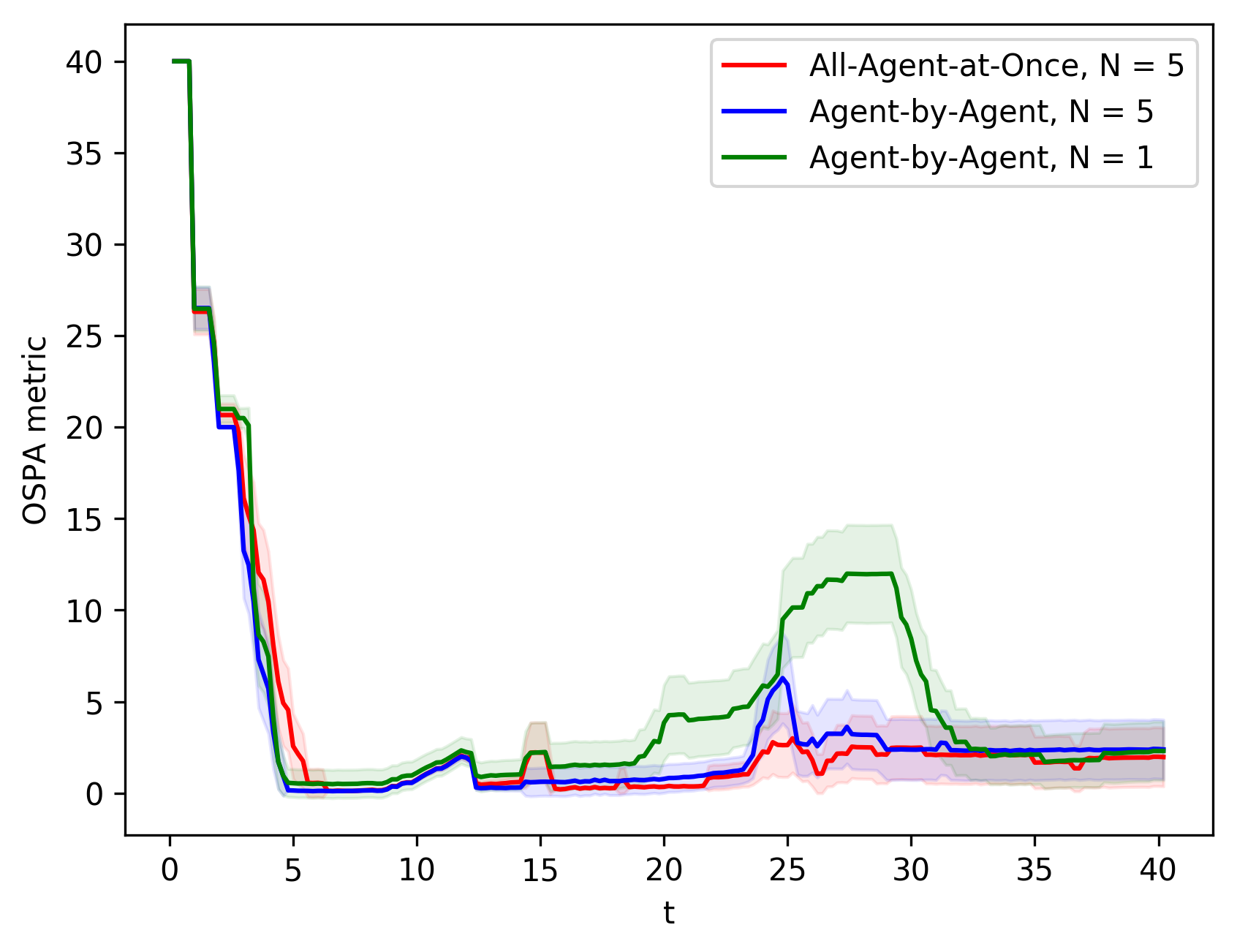}
    \caption{Target tracking performance in OSPA metric, a comparison between the agent-by-agent (distributed) and all-agents-at-once (centralized) rollout policies. 
    Total 50 trials are run in the simulation of 40 decision epochs, with mean and 95\% confidence interval are plotted of the OSPA metric. 
    Parameter of OSPA is with $c = 40.0, p = 2$ following terminology in \cite{schuhmacher2008consistent}.
    This encapsulates both tracking error and target cardinality estimates with respect to ground truth.}
    \label{static_result}
\end{figure}

A typical control result of \textit{agent-by-agent} rollout is represented in Fig. \ref{trajectory}.
The cooperative behavior is observed between the UAV 1 and 2 when Target 1 moves through the bottom green occlusion (between $t = 5$ and $t = 15$): UAV 1 moves to bottom left to continue tracking Target 2 while UAV 2 starts tracking Target 1.
A similar behavior happens again between UAV 2 and 3 when Target 1 moves through the top occlusion area (between $t = 20$ and $t = 30$). 
UAV 2 maintains Target 3 and Target 4 but UAV 3, which initially kept track of Target 4, leaves Target 4 to be tracked by UAV 2 and transitions focus to the track of Target 1.

The performance of the target tracking task is measured by the Optimal Subpattern Assignment (OSPA) \cite{schuhmacher2008consistent} metrics in Fig. \ref{static_result}. 
These results are all generated based on the scenario depicted in Fig. \ref{scenario-setup}.  
Specifically, three algorithms are investigated, an agent-by-agent greedy algorithm ($N=1$) which explicitly optimizes the one-step belief-state reward \eqref{belief-reward} and the two rollout policies, namely all-agents-at-once and agent-by-agent.  
Two major conclusions can be drawn from Fig. \ref{static_result}.  
First the reader can observe from  $t=15$ to $t=35$ the performance of the greedy algorithm suffers. 
This anomaly arises from the myopic policies lack of consideration for future impacts from current action choices and is especially apparent when UAVs need to perform target hand-offs, changing focus from one set of targets to another.  
Both of the rollout algorithms with extended horizons, $N=5$, account for the future trajectory shifts of the targets and the future movements of the cooperative fleet.  
The second realization is the minimal performance differentiation between the two rollout algorithms.
This result shows that the sequential distributed control decision paradigm is a viable replacement for the centralized control.
More specifically this demonstrates that the advantages of the control space dimension reduction from exponential to linear with respect to number of agents \cite{bertsekas2019multiagent} can be employed in the model with partially observable states, e.g., the task of target tracking. 
In the context of implementation, this also yields the advantage of allowing each agent to compute it's own optimized control, distributing the computational load across all agents evenly.

\section{Conclusion}
\label{section_conclusion}

A multiagent (multi-sensor) target tracking problem is formulated as a POMDP.  
The control policy optimization, based on receding horizon rollout was implemented on an agent-by-agent basis.  
This sequential decision making augmentation, multiagent rollout introduced in \cite{bertsekas2019multiagent}, is compared to a standard, all-agents-at-once, rollout policy generation. 
The sequential implementation showed similar performance when implemented in a distributed tracking system where information maintained on a per agent basis and assimilated across the system through consensus-based JPDA.  

With these initial results in hand further investigations will include applications considering limited communication range and/or bandwidth and extending the heterogeneity level of the agents. 
Beyond viable scenario extensions for real-world considerations, the simulations herein have shown a lackluster in computational efficiency which leads to further extension of this sequential methodology either through similar avenues of reinforcement learning alluded to  in \cite{bertsekas2019multiagent} or though approximate dynamic programming techniques which follow the vein of rollout and receding horizon control.


\bibliographystyle{ieeetr}
\bibliography{reference}

\end{document}